# Incorporation of surface correction and anharmonic correction to vacancy formation energy for nickel and copper: bridging the gap between Density Functional Theory and experiment

Prithwish K. Nandi[*] and M. C. Valsakumar

*Materials Physics Division, Indira Gandhi Centre for Atomic Research,
Kalpakkam, Tamil Nadu, India, Pin – 603 102*

[*]*Corresponding author: nandi.prithwish@gmail.com*

## ABSTRACT

Density functional theory (DFT) has been used to estimate vacancy formation enthalpy ($H_f^v$) for a few transition metals like nickel (Ni) and copper (Cu). It is shown that, for these metals, $H_f^v$ is underestimated considerably by DFT. The aim of the present work is to bridge the gap between the estimates made by DFT calculations and experiments. The sources of this discrepancy are identified as to be related to the opening up of the surface like region surrounding the vacancy, and the temperature induced anharmonic contribution. The surface related correction to $H_f^v$ has been estimated by a jellium based model originally proposed by Mattsson *et al.* [Phys. Rev. B 73 (2006) 195123] and subsequently modified by Nandi *et al.* [J. Phys.:Cond. Matt. 22 (2010) 345501]. In this paper, we have estimated the temperature induced anharmonic contribution to $H_f^v$ using density functional perturbation theory. Finally, it is shown that incorporation of both the surface correction and anharmonic correction to $H_f^v$, results in a better agreement with the experimental data.

Keywords: DFT, Density Functional Perturbation Theory, vacancy formation energy, anharmonic correction, surface correction

# 1. Introduction:

In our earlier work [1], we have shown that the vacancy formation energy ($E_f^v$) as calculated using density functional theory (DFT) [2] for a few 3d-transition metals differs significantly from their corresponding experimental values. Such discrepancy is related to the opening up of an electronic surface like region formed as a result of a steep variation of electronic density surrounding the vacant site [1,3-5]. The Kohn - Sham single particle wavefunction makes a transition from oscillatory to a decaying type at such surfaces [3]. This essentially introduces an inaccuracy in the estimate of $E_f^v$ using DFT. Mattsson *et al* [4,5] proposed a model to estimate this surface related correction to $E_f^v$ and applied the model successfully for Pt, Pd and Mo. We, in our earlier work [1], not only showed that the Mattsson's model [5] does not work well for 3d-transitions metals like Ni, Fe and Cr, but also proposed a necessary modification in choosing the parameters of the model. From those calculations, we have shown that though the addition of the surface related correction to vacancy formation energy has improved the agreement with the experimental values, still there is a considerable gap between the two – a fact which motivates us to explore further the possible reasons for such a discrepancy.

It is to be remembered here that the DFT calculations correspond to 0 K, while the experiments for estimating $E_f^v$ are performed at much higher temperature. In fact, there are many experimental techniques available for measuring $E_f^v$ among which positron annihilation experiment is a widely accepted method [6]. This is an equilibrium method and is sensitive to low concentration of vacancies (~$10^{-6}$). From a positron annihilation Doppler broadening spectrum, it is possible to obtain the momentum distribution of the electrons with which the positron annihilates. This distribution of electron momenta for a defect-free crystal will be different from the distribution obtained at high temperature at which there is an exponential increase of the concentration of thermally generated vacancies. Hence, it is possible to measure the change in the trapping rate of positron and from this to deduce the vacancy formation enthalpy [7]. Therefore, to make a meaningful comparison between the experimental and DFT data, it is necessary to estimate the effect of temperature on the vacancy formation energy. More precisely, since the systems are studied at constant pressure the appropriate quantity to be studied is the formation enthalpy $H_f^v$. Starting from thermodynamic relations, it can be shown that vacancy formation enthalpy ($H_f^v$) depends on temperature. From the relation $H = U + PV$ relating the enthalpy ($H$) of a system with internal energy ($U$), pressure ($P$) and volume ($V$), it follows that

$$\left(\frac{\partial H_v^f}{\partial T}\right)_P = T\left(\frac{\partial S_v^f}{\partial T}\right)_P, \qquad (1)$$

where $T$ is the temperature and $S_f^v$ is the vacancy formation entropy. We can expect the relaxation of atoms around the vacancy to increase with increase in lattice parameter. Therefore, in a material with positive thermal expansion coefficient, we expect an increase in relaxation and hence increase in entropy with temperature. The above thermodynamic relation (equation - 1) then implies an increase in the vacancy formation enthalpy with temperature [6]. The dilation of metals with temperature is considered to be originating from the anharmonic contribution of atomic vibrations [6] and therefore, the increase of $H_f^v$ with $T$ may be regarded as a consequence of the anharmonic effects of temperature.

Various methods can be found in the literature for studying the effect of T on $H_f^v$. Foiles has used Monte Carlo simulations and showed that $H_f^v$ increases with T [8]. Later, Megcheche et al. [9] have studied the same using an empirical approach. In this method, they have calculated the vacancy formation enthalpy ($H_f^v$) as a function of lattice parameter. All of these calculations were done without explicitly incorporating the effect of T. Instead, the role of T was introduced via the dependence of lattice parameter on T. There are various methods available to give correspondence between the lattice parameter and the temperature. For example, thermal expansions of Ni and Cu were calculated by Lu *et al.*[10] using the Calphad approach and obtained a reasonable agreement with the experimental data. The analytical expression to describe the dependence of the thermal expansion coefficient ($\alpha$) on $T$ was also proposed by Glazkov [11]. However, Megchiche et al. [9] have used the analytical form proposed by Suh. *et al* [12], where the thermal expansion was measured by dilatometry as well as the X-ray diffraction method. The empirical relations to calculate the value of lattice parameter at a given temperature (T), $L(T)$, is given by:

$$L(T) = L(293)[1 + A_1(T - 293) + A_2(T - 293)^2] \qquad (2)$$

In this study also, they showed that $H_f^v$ increases with T.

In the present calculations, however, we have made an attempt to understand the effect of T on $H_f^v$ using density functional perturbation theory in addition to the calculation of surface related correction to the vacancy formation enthalpy. The detailed methodology for calculating the anharmonic contribution to $H_f^v$ is described in the following section, and the comparisons of thus computed $H_f^v$ values with the experimental data are discussed in Section 3.

## 2. Computational details

We perform the DFT calculations using VASP [13 - 15] (Vienna *Ab initio* Simulation Package) code, using plane-wave basis set. In the present calculations, we use projector augmented wave [16] (PAW) formalism based pseudopotentials (PPs) and PBE [17] exchange-correlation (XC) functional. All the PPs are taken from the VASP PP library. We take great care in ensuring convergence of all the results with respect to system size, basis sets and k-points as discussed in the Appendix. All the calculations done here are based on supercell approach. We perform the calculations with various supercell sizes to study the dependence of the results on the system sizes. We find that $5 \times 5 \times 5$ supercell for both *fcc* Ni and *fcc* Cu (125 atoms) provide convergence of the total energy per atom to better than $10^{-3}$ eV. In all these calculations, we have allowed the ionic positions, volume and shape of the supercell to relax. The relaxation of atomic positions is done with the conjugate gradient method. This minimization process is terminated when the force acting on each atom is less than $10^{-5}$ eV/Å. We perform spin polarized calculations for Ni. The common settings of DFT calculations for Ni and Cu are summarized in the Appendix.

2.1 Surface self-energy corrections

The methodology for calculating the surface self-energy contribution to $H_f^v$ is described in our earlier work [1]. In this work also we have exactly followed the same method.

## 2.2 Anharmonic contribution to $H_f^v$

For both of these metals, the values of $H_f^v$ were calculated at five different lattice parameters ($a_{lat}^T$) using DFT and lattice dilation was defined as $\Delta a/a_{lat}^0$ where, $\Delta a = a_{lat}^T - a_{lat}^0$ and $a_{lat}^0 = equlibrium\ lattice\ constant$. This data was fitted with a second order polynomial of the form:

$$H_f^v = A\left(\frac{\Delta a}{a_{lat}^0}\right)^2 + B\left(\frac{\Delta a}{a_{lat}^0}\right) + C, \qquad (3)$$

where $A, B\ and\ C$ are the fitting parameters. In Fig. 1(a) and 1(b), we have plotted the values of $H_f^v$ vs. $\Delta a/a_{lat}^0$ along with the fitted curve, for both Ni and Cu respectively. It can be seen from this data that the values of $H_f^v$ increases with the values of $\Delta a/a_{lat}^0$. Now the main task is to estimate $\frac{\Delta a}{a_{lat}^0}$ at a desired temperature. In this work, we have calculated the value of lattice parameter at a temperature, T using *ab initio* method where anharmonic effects are incorporated indirectly via quasiharmonic approximation (QHA) [19,20]. In order to calculate thermal expansion at ambient pressure, we define a function $X$ such that,

$$X(V; P, T) = U(V) + F_{phonon}(V; T) + PV \qquad (4)$$

where, $P, T$ and $V$ represent pressure, temperature and volume respectively. $U(V)$ is taken to be the electronic total energy as calculated by VASP. $F_{phonon}(V; T)$ is the Helmholtz free energy of the phonon system and it has been calculated here using the PHONOPY [20] package. This is a computer code which is used to set up the dynamical matrix using the *ab initio* force constant data generated by VASP and to calculate the phonon frequencies and the thermal properties. The $PV$ term represents the work done on the system, but at ambient pressure this term can be neglected as it is small. The minimum of $X(V;P,T)$ at a particular volume $V$, is defined as the Gibb's free energy, $G$ and it describes the phase stability:

$$G(P;T) = min_V X(V;P,T) \qquad (5)$$

In the present work, we have optimized the crystal structures for eleven volumes equally spaced on either side of the equilibrium volume. The phonon energies at each volume were calculated for a wide range of temperature (0 - 1000 K for Cu and 0 - 1600 K for Ni) using PHONOPY. Finally, at each temperature, T, the equilibrium volume was determined by fitting the $X(V;P,T)$ vs. $V$ data to the Murnaghan equation of state [21]:

$$E(V) = E_0 + \frac{B_0 V}{B_0'}\left(\frac{(V_0/V)^{B_0'}}{B_0'-1} + 1\right) - \frac{B_0 V_0}{B_0'-1} \qquad (6)$$

Thus by knowing the equilibrium volume at different temperatures, linear thermal expansion, $\Delta a_{lat}/a_{lat}$ (in %) can be calculated. Finally, we calculated $H_f^v$ as a function of $\Delta a_{lat}/a_{lat}$ using the equation - (3).

## 3. Results and Discussions

The values of the vacancy formation energy for Ni and Cu have been calculated using the following equation [1]:

$$E_f^v = E(N-1,1) - \frac{N-1}{N}E(N,0) \qquad (7)$$

Here, E(N,0) represents total energy of the perfect system with N atoms of the supercell and E(N-1,1) is the energy of the system when one of the atoms is replaced by a vacancy.

In Table 1, we have shown the calculated values of vacancy formation enthalpy for Ni and Cu and also compared these values with the experimental data as collected from the literature. Please note that DFT underestimates $E_f^v$ by ~20% and ~23% respectively for Ni and Cu. We should also mention here that these discrepancies are unrelated to lattice relaxations, since in all our calculations, we have incorporated full lattice relaxations.

In order to reduce this discrepancy, first, we have incorporated the surface related correction to $H_f^v$ following the prescription of our earlier work [1]. We have seen that for both Ni and Cu, the magnitude of this surface related correction is similar (0.16 eV). It is important to notice here is that even after adding this correction to the DFT value of $H_f^v$, the agreement with experiment has not yet been achieved, though the magnitude of the discrepancy has been reduced. Hence, we go to the next level of calculations.

Since the experiments to estimate $H_f^v$ are done at high temperatures, here, we have made an attempt to estimate the effect of $T$ on $H_f^v$. It is to be mentioned here that the effect of $T$ on $H_f^v$ has been introduced via the dependence of lattice parameter on $T$, instead of explicitly incorporating the effect of $T$ on $H_f^v$. In order to calculate, $a_{lat}(T)$ as a function of $T$, we have followed the methodology described in section 2. The magnitudes of $a_{lat}(T)$ at different temperatures have been obtained from the fitting parameters of equation 6. In Figs. 2(a) and 2(b), we have plotted the free energy, $X(P; V, T)$ as a function of volume, $V$ at different temperatures, $T$ for Ni and Cu respectively. The volumes ($V_0$) corresponding to the minimum free energy at each $T$ has been shown by the dashed line in each of these two figures. Please note that, for both of these metals, $V_0$ increases with the increment of $T$. Having the knowledge of $V_0$ at different temperatures, it is possible for us to calculate linear thermal expansion, $\Delta a_0/a_0$ (in %) as a function of $T$. In Figs. 3(a) and 3(b), we have plotted the variation of $\Delta a_0/a_0$ with $T$ as obtained from our calculations.

Here, $a_0$ is the lattice parameter at 0 K. We have also shown the corresponding experimental data (by the triangular symbols) as obtained from XRD experiments [12]. Since, the experimental data for the linear expansion is calculated as $\Delta a_{300}/a_{300}$, where, $a_{300}$ is the lattice parameter at $T$ = 300 K, we have scaled the computed values of $\Delta a_0/a_0$ with a constant factor $a_0/a_{300}$ as obtained from our calculations. These plots show an overall agreement of the computed data with the experimental values, though deviations are seen at high temperature. We should mention here that, in QHA, the full Hamiltonian is replaced by a harmonic expansion about the equilibrium positions at a given volume [8]. Such harmonic approximations may not be valid when the temperature is very high [8].

Having the knowledge of $\Delta a_0/a_0$ (in %) as a function of $T$, we calculated $H_f^v$ at various temperatures using equation 3. In Figs. 4(a) and 4(b), we have shown the variation of $H_f^v$ with $T$ for Ni and Cu respectively. The horizontal dashed line at the bottom of each figure shows the values of $H_f^v$ as obtained from our DFT calculations using PAW PBE. The horizontal dotted lines (in green) represent the available experimental data for $H_f^v$. This clearly shows that there is a considerable mismatch between the data as obtained from experiments and the DFT calculations. To mitigate the gap between the two, first, we have added the temperature induced anharmonic correction to the DFT values of $H_f^v$ and this is shown by the blue line in these two figures. Finally, the surface related correction was also added to this data. The curve in red, in fact, shows the dependence of $H_f^v$ on $T$ after introducing both the surface related correction and the temperature induced anharmonic contribution to $H_f^v$ as obtained from DFT calculations.

The comparison of the corrected values of $H_f^v$ with the experimental data is not straightforward. It is seen from these figures that many estimates of $H_f^v$ are available in the literature, though all the data quoted here are obtained using positron annihilation experiments. In fact, there are many technical details regarding the measurements of vacancy formation enthalpy and detailed descriptions of these are well beyond the scope of the present work. However, to make a brief comment on the different estimates of $H_f^v$, it important to give a glimpse of these works. Nanao et al [22] measured the peak count rate as a function of $T$ above room temperature in the solid and liquid phases of Cu and in the solid phase of Ni. They analysed the data numerically using the trapping model. For Ni, they quoted two estimates: 1.74±0.06 eV assuming the peak count rate due to the positron annihilating in the perfect region is linear with $T$ and 1.65±0.06 eV assuming the peak count rate to be quadratic with $T$. Similarly for Cu also, they provided two estimates: 1.28±0.04 eV and 1.21±0.04 eV. Smadskjaer et al [7] estimated the value of $H_f^v$ as 1.8±0.01 eV for Ni using the Doppler broadening (DB) technique. Another estimate of $H_f^v$ for Ni was made by Campbell et al [23] who also used the Doppler broadened annihilation gamma ray line shape measured between room temperature and the melting point. Their estimate for $H_f^v$ is 1.73 eV assuming linear temperature dependence of the untrapped positron line shape. Similarly, for Cu, two different estimates were made by Rice-Evans et al [24]. They studied the annihilation of positrons in Cu as a function of temperature and obtained the value as 1.32 eV using the data as collected using a low energy photon Ge(Li) detector above 900 K, whereas the other estimate is 1.26±0.07 eV when the thermal expansion was taken into account.

From the above discussions, it is clear that the experimental values of $H_f^v$ are very much sensitive to the accuracy of the measurements, the selection of a temperature regime of the obtained data and the usage of appropriate of numerical model for analyzing these data. Moreover, these works did not describe the variation of $H_f^v$ with temperature. Therefore, it is difficult for us to compare our data directly with experiments. But one thing we should notice is that, for Ni, the calculated values of $H_f^v$ in the temperature range 1250 – 1450 K) match well with the experimental estimates. It is to be mentioned here that in this temperature range, the positron line shape appears to be linear with $T$, as seen from the work of Smadskjaer et al [7]. On the other hand, the experimental value of $H_f^v$ for Cu is 1.32 eV [22] (as obtained using data for T > 900 K) is also consistent with our calculation.

One final comment we should mention here is that in this calculation we have only introduced the quasi-harmonic contributions of T. The full anharmonic contribution can also be calculated following the method as outlined by Grabowsky et al [25], but the procedure is computationally very much expensive. However, the strength of the present calculation is in pointing the sources of discrepancies between the DFT data and the experimental values of $H_f^v$. Our calculations also provide the trend of the variation of $H_f^v$ with $T$ and we should expect to see experiments exploring the same in near future, so that a direct experimental validation of the present calculated data can be made.

## 4. Conclusions

This paper deals with the issues related to the accurate estimation of vacancy formation energy for Ni and Cu using DFT. The detailed DFT study of bulk properties like equilibrium lattice parameter and bulk modulus for these 3d-transition metals have been carried out using PBE exchange – correlation functional and PAW pseudopotential. Our results demonstrate that DFT calculations make inaccurate estimate for vacancy formation energy. Therefore, we conclude that even the so-called simple problem of calculating vacancy formation energy is not straightforward. Attempts have been made to resolve this issue by incorporating surface intrinsic energy corrections to $E_f^v$ using a jellium based model originally developed by Mattsson *et al.* [4,5] and subsequently modified by Nandi et al. [1] and also adding the anharmonic contribution of temperature to $E_f^v$ following the density functional perturbation theory (DFPT). It has been shown that incorporation of both the surface correction and anharmonic correction to $E_f^v$, improves the agreement of the DFT values with the experimental data.

## Acknowledgements

One of the authors (PKN) wants sincerely to acknowledge Dr. C. S. Sundar and Ms. Gurpreet Kaur for some useful discussions.

## Appendix

*(a) Settings for calculating the vacancy formation energy:*

Common settings for all Ni calculations: plane wave cutoff is ~337.0 eV for PAW PBE which is more than the recommended cutoff energy (ENMAX) of 269.5 eV. Augmentation used ~545 eV. In all calculations for Ni the numbers of k-points used are 4 × 4 × 4 in the Monkhorst-Pack scheme [18]. This gives the convergence of ~$10^{-5}$ eV for the total energy per atom.

Common settings for all Cu calculations: plane wave cutoffs are 355.2 eV for PAW PBE, which is more than the recommended cutoff energy (ENMAX) of 273.2 eV and the value augmentation used is 516.5 eV. In all calculations for Cu the numbers of k-points used are $4 \times 4 \times 4$ in the Monkhorst-Pack scheme [18]. This gives the convergence of $\sim 10^{-5}$ eV for the total energy per atom.

For all calculations mentioned above the energy tolerance for electronic iterations are $10^{-6}$ eV and Fermi smearing value is 0.2 eV. All the calculations are performed with "PRECISION = HIGH" in the INCAR files.

*(b) Settings for calculating the dynamical matrix:*

For Ni: The k - point mesh for a $2 \times 2 \times 2$ supercell (32 atoms) is $12 \times 12 \times 12$. The convergence criterion for the electronic steps is $10^{-9}$ eV. IBRION = 8 is used for all the calculations.

For Cu: The k - point mesh for a $2 \times 2 \times 2$ supercell (32 atoms) is $18 \times 18 \times 18$. The convergence criterion for the electronic steps is $10^{-9}$ eV. IBRION = 8 is used for all the calculations.

**Table 1.** The computed DFT values of equilibrium lattice parameters, bulk moduli and vacancy formation enthalpy for Ni and Cu. The values are calculated using PAW PBE. The computed values are compared with experimental values.

| Metal | $a_{lat}$ (Å) | | $B_0$ (GPa) | | $H_f^v$ (eV) | |
|---|---|---|---|---|---|---|
| | DFT | Expt. [26] | DFT | Expt. [26] | DFT | Expt. |
| Ni | 3.523 | 3.524 | 193.64 | 180 | 1.42 | 1.8±0.01 [7] <br> 1.74±0.06 [22] <br> 1.65±0.06 [22] <br> 1.73 [23] |
| Cu | 3.651 | 3.615 | 131.99 | 140 | 1.05 | 1.32 [24] <br> 1.26±0.07 [24] <br> 1.28±0.04 [22] <br> 1.21±0.04 [22] |

# Figure Captions

**Fig. 1** Vacancy formation enthalpy vs. dilation ($\Delta a/a_{lat}^0$) : (a) for Ni, (b) for Cu. DFT data have been shown as red filled circles and the line (in black) represents the curve fitted to the DFT data. The data was fitted with a second order polynomial of degree two: $H_f^v = A\left(\frac{\Delta a}{a_{lat}^0}\right)^2 + B\left(\frac{\Delta a}{a_{lat}^0}\right) + C$

**Fig. 2** Variation of Helmholtz free energy with volume at different temperatures: (a) for Ni and (b) for Cu. The dashed curve shows the equilibrium volumes at various temperatures. Note that, for both Ni and Cu, the lattice dilates with the increase of temperature.

**Fig. 3.** Linear thermal expansion (in %) is plotted as a function of temperature, T: (a) for Ni and (b) for Cu. The triangles show the experimental points obtained from X-ray diffraction by Suh et al. [12]. The calculated data points are seen to be in good agreement with the experiments. Note that, for both Ni and Cu, the lattice dilates with increase of temperature.

**Fig. 4.** This plot shows how the vacancy formation enthalpy ($H_f^v$) varies with temperature, T: (a) for Ni and (b) for Cu. The horizontal dashed line at the bottom of each figure shows the value as calculated by DFT using PAW PBE. The green horizontal lines show the experimental values obtained from positron experiments. This shows clearly that there is a large discrepancy between the experiment and DFT calculations. Efforts have been made to mitigate this discrepancy by introducing surface related correction as outlined by Nandi et al [1] and effect of temperature on the vacancy formation enthalpy using DFPT. The blue curve shows the $H_f^v$ values after adding the anharmonic contribution of T to $H_f^v$, which is still differing significantly from the experimental data. The red curve shows the final calculated values of $H_f^v$ after adding the surface related corrections. Please note that after introducing the corrections, the calculated values for $H_f^v$ agree well with the experiment.

# Table Captions

**Table 1.** The computed DFT values of equilibrium lattice parameters, bulk moduli and vacancy formation enthalpy for Ni and Cu. The values are calculated using PAW PBE. The computed values are compared with experimental values.

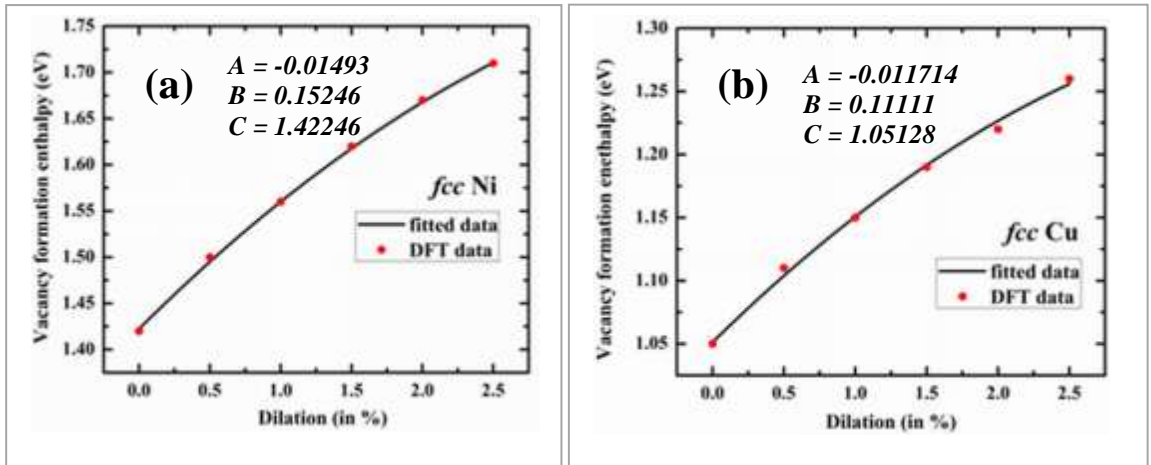

**Fig. 1** Vacancy formation enthalpy vs. dilation ($\Delta a/a_{lat}^0$) : (a) for Ni, (b) for Cu. DFT data have been shown as red filled circles and the line (in black) represents the curve fitted to the DFT data. The data was fitted with a second order polynomial of degree two:

$$H_f^v = A\left(\frac{\Delta a}{a_{lat}^0}\right)^2 + B\left(\frac{\Delta a}{a_{lat}^0}\right) + C$$

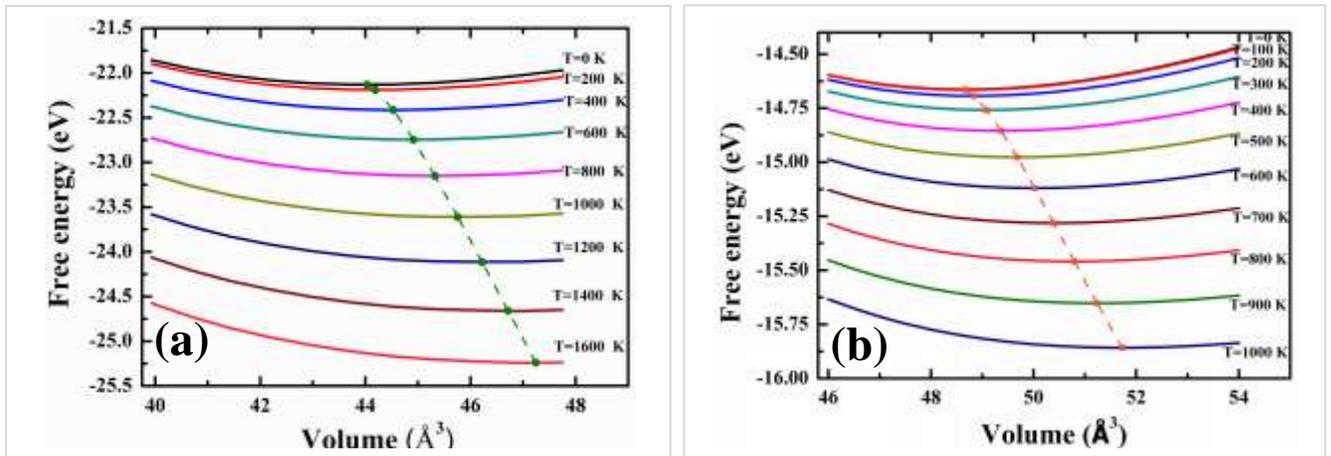

**Fig. 2** Variation of Helmholtz free energy with volume at different temperatures: (a) for Ni and (b) for Cu. The dashed curve shows the equilibrium volumes at various temperatures. Note that, for both Ni and Cu, the lattice dilates with the increase of temperature.

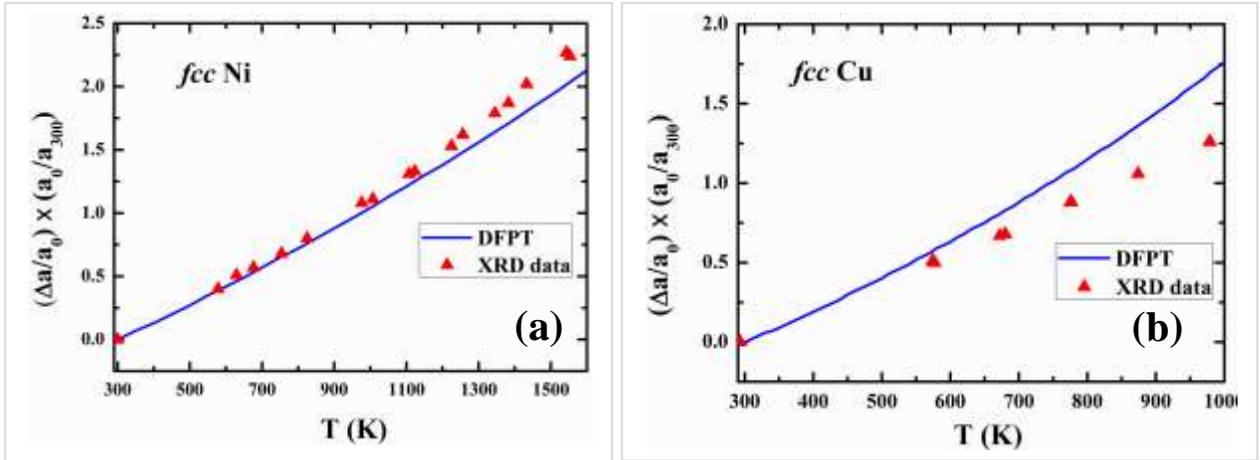

**Fig. 3.** Linear thermal expansion (in %) is plotted as a function of temperature, T: (a) for Ni and (b) for Cu. The triangles show the experimental points obtained from X-ray diffraction by Suh et al. [12]. The calculated data points are seen to be in good agreement with the experiments. Note that, for both Ni and Cu, the lattice dilates with increase of temperature.

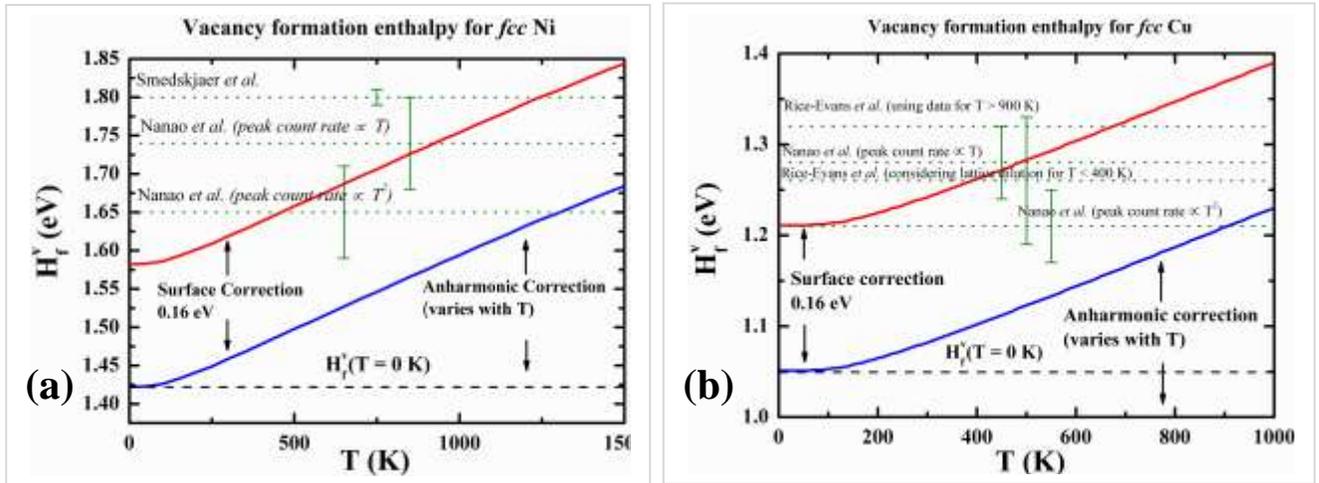

**Fig. 4.** This plot shows how the vacancy formation enthalpy ($H_f^v$) varies with temperature, T: (a) for Ni and (b) for Cu. The horizontal dashed line at the bottom of each figure shows the value as calculated by DFT using PAW PBE. The green horizontal lines show the experimental values obtained from positron experiments. This shows clearly that there is a large discrepancy between the experiment and DFT calculations. Efforts have been made to mitigate this discrepancy by introducing surface related correction as outlined by Nandi et al [1] and effect of temperature on the vacancy formation enthalpy using DFPT. The blue curve shows the $H_f^v$ values after adding the anharmonic contribution of T to $H_f^v$, which is still differing significantly from the experimental data. The red curve shows the final calculated values of $H_f^v$ after adding the surface related corrections. Please note that after introducing the corrections, the calculated values for $H_f^v$ agree well with the experiment.